\documentclass[aps,pre,twocolumn]{revtex4-2}

\usepackage{graphicx}
\usepackage{amsmath}
\usepackage{bm}
\DeclareMathOperator{\Tr}{Tr}

\begin{document}

\title{Modulated phases of nematic liquid crystals induced by tetrahedral order}
\author{Michely P. Rosseto}
\affiliation{Departamento de F\'isica, Universidade Estadual de Maring\'a, Maring\'a, Paran\'a 5790-87020-900, Brazil}
\author{Jonathan V. Selinger}
\affiliation{Department of Physics, Advanced Materials and Liquid Crystal Institute, Kent State University, Kent, Ohio 44242, USA}
\date{December 30, 2021}
\begin{abstract}
Recent theoretical research has developed a general framework to understand director deformations and modulated phases in nematic liquid crystals.  In this framework, there are four fundamental director deformation modes:  twist, bend, splay, and a fourth mode $\bm{\Delta}$ related to saddle-splay.  The first three of these modes are known to induce modulated phases.  Here, we consider modulated phases induced by the fourth mode.  We develop a theory for tetrahedral order in liquid crystals, and show that it couples to the $\bm{\Delta}$ mode of director deformation.  Because of geometric frustration, the $\bm{\Delta}$ mode cannot fill space by itself, but rather must be accompanied by twist or splay.  Hence, it may induce a spontaneous cholesteric phase, with either handedness, or a splay nematic phase.
\end{abstract}

\maketitle

\section{Introduction}
\label{introduction}

In nematic liquid crystals, the molecules align along a local axis, called the director $\hat{\bm{n}}$.  In the simplest nematic phase, the director tends to be uniform.  However, in modulated versions of the nematic phase, the director field $\hat{\bm{n}}(\bm{r})$ varies as a function of position $\bm{r}$ in a periodic structure.  The most common example of a modulated nematic phase is the cholesteric phase, in which the director field forms a helix.
More complex examples are blue phases, which have an array of tubes with double twist in the director field, separated by disclination lines, arranged in cubic lattices.  Many other modulated structures have been predicted theoretically, and some of them have been reported experimentally over the past decade, including the twist-bend nematic phase~\cite{Meyer1976,Dozov2001,Chen2013,Borshch2013,Shamid2013,Barbero2015} and the splay nematic phase~\cite{Mertelj2018,Mandle2019,Connor2020,Sebastian2020,Copic2020,Rosseto2020}.

In a recent review article~\cite{Selinger2022}, our group proposed a unified framework to understand all of the modulated nematic phases.  This approach is based on two general principles:

First, we consider the four fundamental deformation modes of the nematic director field---twist, bend, splay, and a less-well-known fourth mode called $\bm{\Delta}$, related to saddle-splay---which have been identified in recent theoretical research~\cite{Machon2016,Selinger2018}.
Each of these director deformation modes can induce some type of molecular order, in addition to the standard nematic orientational order:  twist induces chirality, bend induces polar order perpendicular to $\hat{\bm{n}}$, splay induces polar order parallel to $\hat{\bm{n}}$, and $\bm{\Delta}$ induces tetrahedral order.  The concept of tetrahedral order has been discussed theoretically~\cite{Lubensky2002,Gaeta2016}.  Conversely, each type of molecular order might form spontaneously, to make a phase with nematic order and a small amount of extra order.  In such a phase, the extra order induces an ideal local structure with the corresponding director deformation.  The most common case is that chirality induces twist, but the other three cases are also possible.

Second, we consider the concept of geometric frustration, which describes an ideal local structure that cannot fill three-dimensional (3D) Euclidean space because of geometric constraints.  This concept is widely used in recent research on solid materials~\cite{Grason2016,Meiri2021}.  In the context of liquid crystals, an ideal local director deformation is generally frustrated; only a few special combinations of deformation modes can fill space~\cite{Virga2019}.  For that reason, the liquid crystal must form a complex global phase, which may have a combination of favorable and unfavorable deformation modes, or may have a periodic array of defects.

Based on those principles, our review article analyzed several types of modulated phases induced by extra molecular order~\cite{Selinger2022}.  If a liquid crystal has chirality, then the ideal local structure has pure twist (i.e.\ double twist), but this pure twist cannot fill space.  In response to that frustration, the liquid crystal may form a cholesteric phase, with a combination of twist and $\bm{\Delta}$ mode.  Alternatively, it may form a blue phase, with tubes of almost pure twist separated by disclination defects.  By comparison, if a liquid crystal has polar order perpendicular to the director, then the ideal local structure has pure bend of constant magnitude, which also cannot fill space.  In this case, the liquid crystal may form a twist-bend nematic phase, with a combination of bend, twist, and $\bm{\Delta}$ mode.  Likewise, if a liquid crystal has polar order parallel to the director, then the ideal local structure has pure splay of constant magnitude, which again cannot fill space.  This liquid crystal may form a splay nematic phase, which has regions of splay separated by defects, and which includes a large component of $\bm{\Delta}$ mode.

This theoretical analysis leaves one conspicuous open question:  If modulated phases can be induced by three of the four types of extra molecular order, then what about the fourth type?  Suppose that a liquid crystal has spontaneous tetrahedral order, which creates an ideal local structure with the $\bm{\Delta}$ deformation mode.  Will that liquid crystal form a modulated phase?  If so, what is the structure of the phase?

\begin{figure*}
(a)\includegraphics[height=4.5cm]{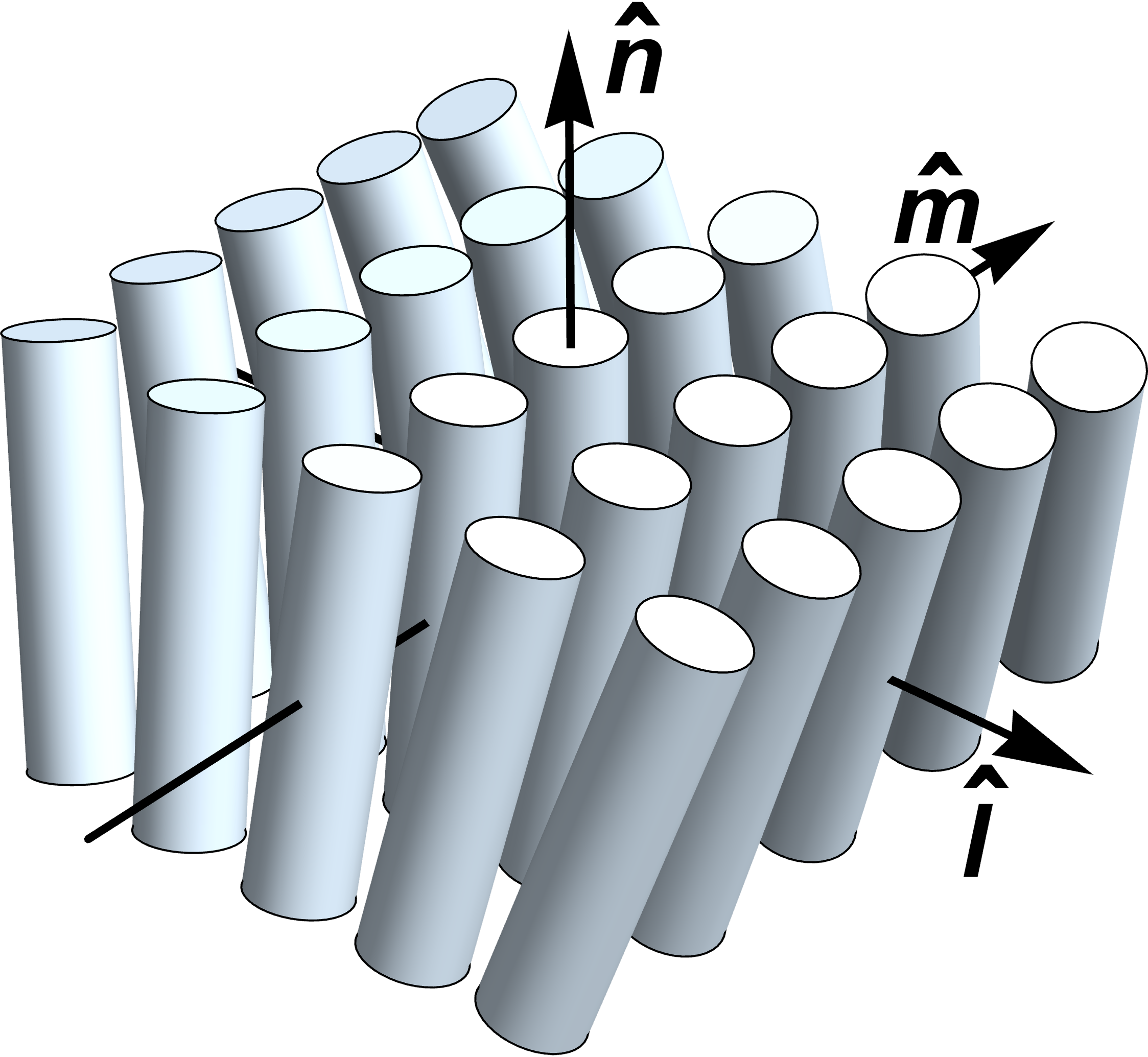}
(b)\includegraphics[height=4.5cm]{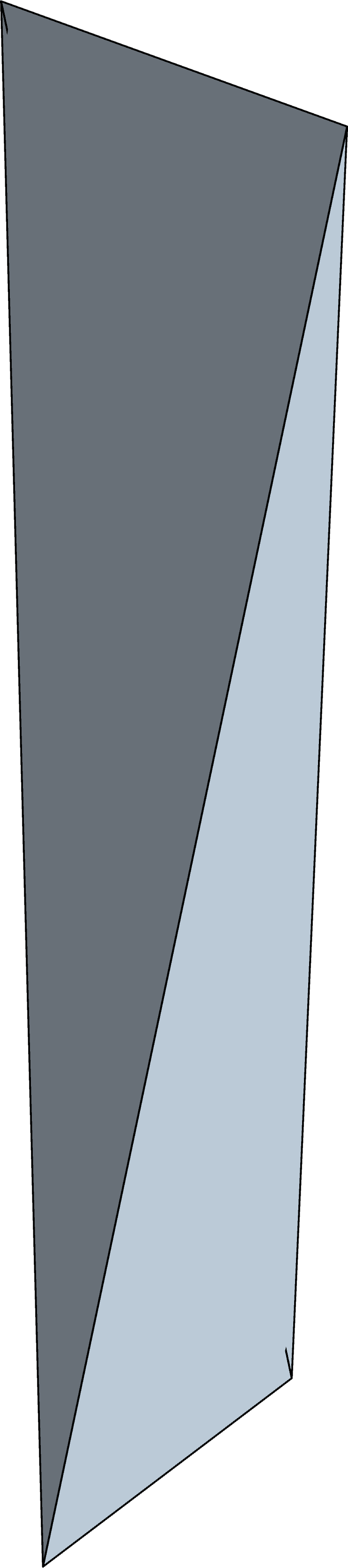}
(c)\includegraphics[height=4.5cm]{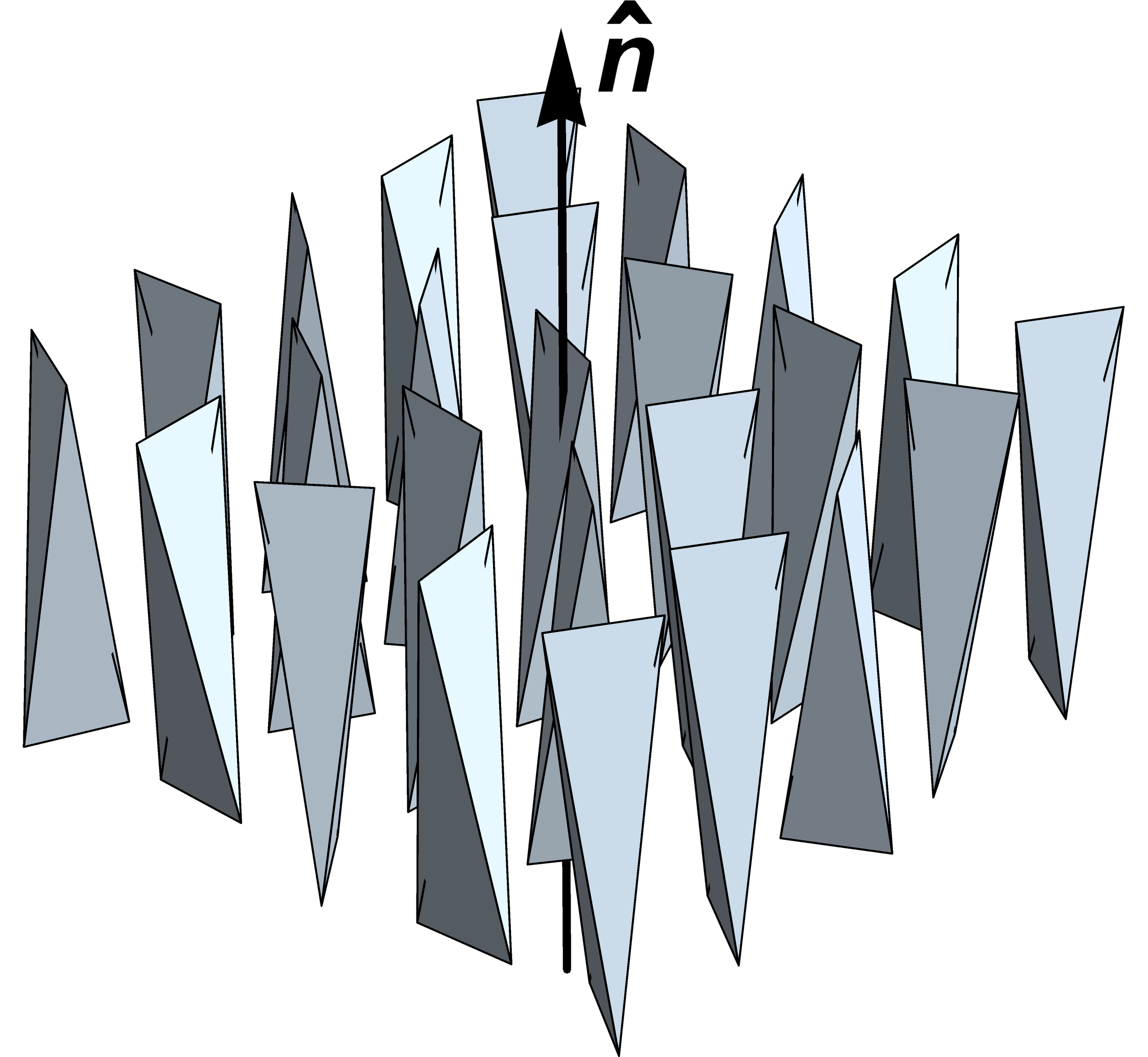}
(d)\includegraphics[height=4.5cm]{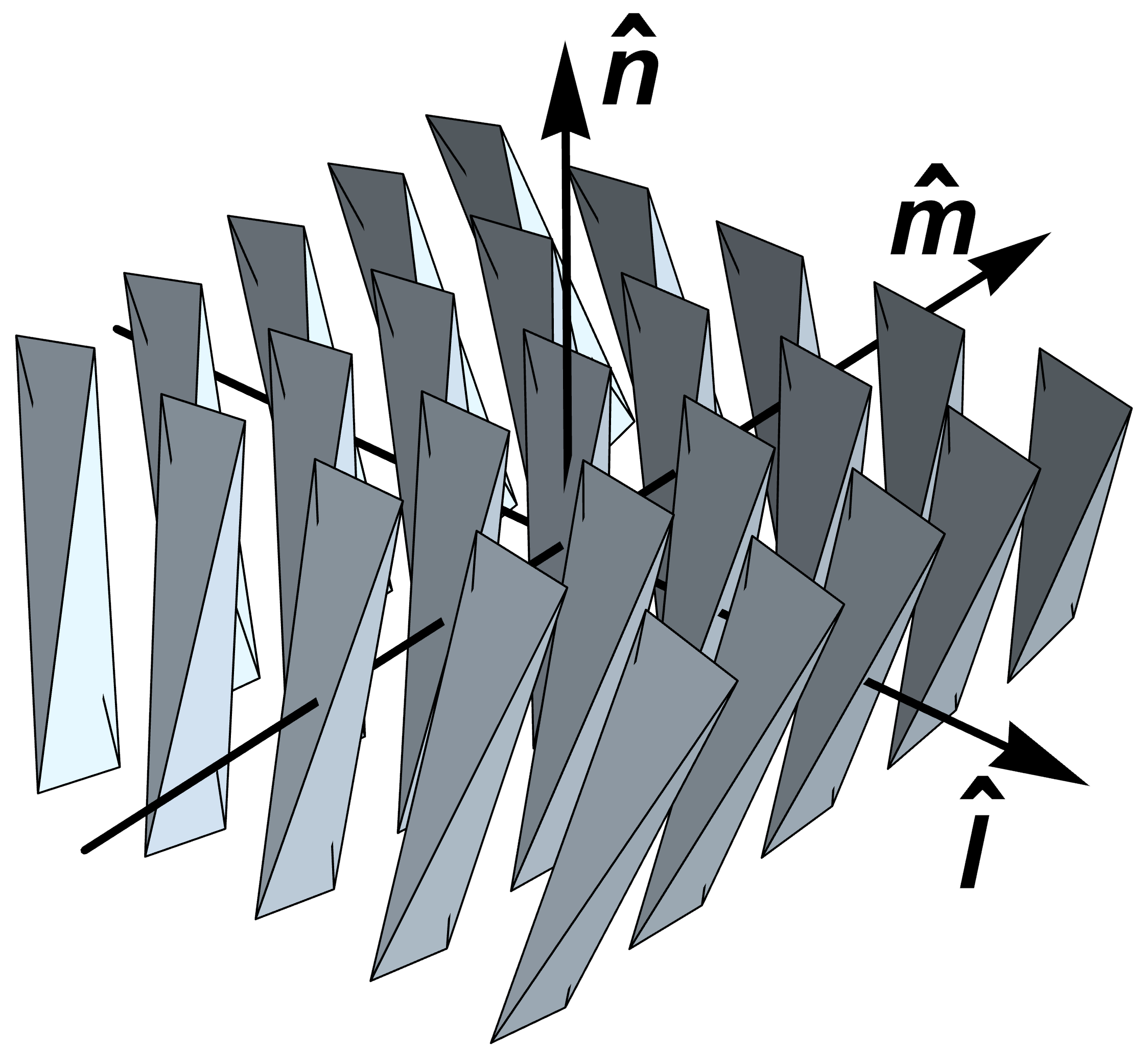}
\caption{(a)~Schematic illustration of the $\bm{\Delta}$ mode of director deformation.  (b)~Example of a distorted tetrahedral particle.  Molecules of this shape should be highly compatible with the $\bm{\Delta}$ mode.  (c)~Uniaxial nematic phase of distorted tetrahedra \emph{without} tetrahedral order, because the molecules have random orientations in the plane perpendicular to $\hat{\bm{n}}$.  (d)~Nematic phase of distorted tetrahedra \emph{with} tetrahedral order, which is aligned with the $\bm{\Delta}$ deformation of the director field.}
\label{fig:tetrahedra}
\end{figure*}

The purpose of the current paper is to answer that question.  In Sec.~\ref{sec:tetrahedra}, we visualize tetrahedral order and show how it is coupled with the $\bm{\Delta}$ mode of director deformation.  In Sec.~\ref{sec:phases}, we consider modulated structures that might form in a liquid crystal with tetrahedral order.  One possibility is a cholesteric phase, which would combine the favorable $\bm{\Delta}$ mode with the unfavorable twist mode.  This version of the cholesteric phase could be either right- or left-handed; it would spontaneously break reflection symmetry.  Another possibility would have the same 1D or 2D structure as the splay nematic phase, but it would be stabilized by $\bm{\Delta}$ mode rather than by splay.  We compare the free energies of these structures and derive a phase diagram.  In Sec.~IV, we discuss the significance of these results for understanding modulated phases in liquid crystals.

\section{$\bm{\Delta}$ mode and tetrahedral order}
\label{sec:tetrahedra}

In this section, we describe the $\bm{\Delta}$ mode of director deformation, and show how it is coupled with tetrahedral order, using an argument analogous to the theory of flexoelectricity.  We present the argument first visually and then mathematically.

In his classic article on flexoelectricity~\cite{Meyer1969}, Meyer showed that the bend mode has the same symmetry as a banana, and hence it is highly compatible with banana-shaped molecules.  If a bend deformation is applied to a uniaxial nematic phase of banana-shaped molecules (or molecules with a banana-like component to the shape), it aligns the orientational distribution of the molecules, and hence induces polar order perpendicular to the director.  This polar order is observed as the bend flexoelectric effect.  Similarly, the splay mode has the same symmetry as a pear, and hence it is highly compatible with pear-shaped molecules.  If a splay deformation is applied to a nematic phase of pear-shaped molecules (or molecules with a pear-like component to the shape), it aligns the pears up or down along the director, and hence induces polar order parallel to the director.  That polar order is observed as the splay flexoelectric effect.

Let us apply the same argument to the $\bm{\Delta}$ mode of director deformation.  The $\bm{\Delta}$ mode has the structure shown in Fig.~\ref{fig:tetrahedra}(a).  The director $\hat{\bm{n}}$ tilts outward along one axis $\hat{\bm{l}}$, and inward along the orthogonal axis $\hat{\bm{m}}$, in the plane perpendicular to $\hat{\bm{n}}$.  This deformation was first defined by Machon and Alexander~\cite{Machon2016}, who called it ``anisotropic orthogonal gradients of $\hat{\bm{n}}$.''  It is related to the so-called surface elastic mode of saddle-splay, but we have argued that it can more usefully be regarded as a bulk elastic mode~\cite{Selinger2018}.

The $\bm{\Delta}$ mode has the same symmetry as a distorted tetrahedron, shown in Fig.~\ref{fig:tetrahedra}(b).  This shape is extended along one axis, and it tilts outward along one axis and inward along the orthogonal axis.  Following the flexoelectric analogy, the $\bm{\Delta}$ mode should be highly compatible with molecules shaped like this distorted tetrahedron, or at least with a tetrahedral component to the shape.  For that reason, we have suggested that the $\bm{\Delta}$ mode might be called ``tetrahedral splay''~\cite{Selinger2022}.

A uniaxial nematic phase of distorted tetrahedra has the structure shown in Fig.~\ref{fig:tetrahedra}(c).  The long axes of the tetrahedra are aligned along the director $\hat{\bm{n}}$, and the transverse axes of the molecules are random.  Because of the randomness, this structure does \emph{not} have tetrahedral order.  If a $\bm{\Delta}$ deformation of the director field is applied to the phase, it aligns the transverse axes of the molecules along the $\hat{\bm{l}}$ and $\hat{\bm{m}}$ directions, as shown in Fig.~\ref{fig:tetrahedra}(d).  That structure now has tetrahedral order, in the plane perpendicular to $\hat{\bm{n}}$.  In the figure, we can see that the combination of $\bm{\Delta}$ deformation and tetrahedral order gives a very efficient packing of the molecules.

The same physical mechanism works in reverse, as in the converse flexoelectric effect.  If a liquid crystal has a spontaneous tendency to form tetrahedral order, this tetrahedral order will induce the $\bm{\Delta}$ deformation.

Now let us express the same argument mathematically.  In elasticity theory, we need to classify all gradients of the director.  References~\cite{Machon2016,Selinger2018} show that the director gradient tensor can be decomposed into four distinct mathematical objects as
\begin{equation}
\partial_i n_j = -n_i B_j + \frac{1}{2}S(\delta_{ij}-n_i n_j) + \frac{1}{2}T\epsilon_{ijk}n_k + \Delta_{ij}.
\end{equation}
Here, $\bm{B}=\hat{\bm{n}}\times(\bm{\nabla}\times\hat{\bm{n}})=-(\hat{\bm{n}}\cdot\bm{\nabla})\hat{\bm{n}}$ is the bend vector, $S=\bm{\nabla}\cdot\hat{\bm{n}}$ is the splay scalar, $T=\hat{\bm{n}}\cdot(\bm{\nabla}\times\hat{\bm{n}})$ is the twist pseudoscalar, and $\Delta_{ij}$ is the remaining component.  Mathematically, $\Delta_{ij}$ is a symmetric, traceless tensor in the plane perpendicular to $\hat{\bm{n}}$.  It has eigenvalues of 0 and $\pm\delta$, with corresponding eigenvectors of $\hat{\bm{n}}$, $\hat{\bm{l}}$, and $\hat{\bm{m}}$.  Hence, it can be written as $\Delta_{ij}=\delta(l_i l_j - m_i m_j)$.

The second-rank tensor $\Delta_{ij}$ is odd in $\hat{\bm{n}}$, and hence it changes sign under the nematic symmetry of $\hat{\bm{n}}\leftrightarrow-\hat{\bm{n}}$.  If we want a uniquely defined physical object, we can construct the third-rank or octupolar tensor $\Delta_{ij}n_k$, which is even in $\hat{\bm{n}}$.  This tensor describes the distorted tetrahedral symmetry of the deformation.  It is not the most general octupolar tensor, but rather has several special features:  the first two legs are traceless, symmetric, and perpendicular to $\hat{\bm{n}}$, while the third leg is parallel to $\hat{\bm{n}}$.

Next, we define a tensor order parameter to represent the tetrahedral order shown in Fig.~\ref{fig:tetrahedra}(d).  This order parameter must also be a third-rank or octupolar tensor $O_{ijk}$, with the same special features as $\Delta_{ij}n_k$.  Based on these features, the octupolar tensor can be written as $O_{ijk}=\Omega_{ij}n_k$, where $\Omega_{ij}$ is a symmetric, traceless tensor in the plane perpendicular to $\hat{\bm{n}}$, and $\Omega_{ij}$ changes sign under the symmetry $\hat{\bm{n}}\leftrightarrow-\hat{\bm{n}}$.  

The free energy density of the liquid crystal is now the sum of three components, $F=F_\text{nem}+F_\times+F_\text{tet}$.  First, there is the Oseen-Frank free energy density associated with director gradients.  As discussed in Refs.~\cite{Selinger2018,Selinger2022}, it can be written as
\begin{align}
F_\text{nem} &= \frac{1}{2} (K_{11} -K_{24}) S^2 + \frac{1}{2} (K_{22}-K_{24}) T^2 \nonumber\\
&\quad + \frac{1}{2} K_{33} |\bm{B}|^2 + K_{24} \Tr(\bm{\Delta}^2),
\label{Fnem}
\end{align}
where $(K_{11} -K_{24})$, $(K_{22}-K_{24})$, $K_{33}$, and $K_{24}$ are the elastic constants for splay, twist, bend, and $\bm{\Delta}$ mode, respectively.  Second, there is a bilinear coupling between director gradients and tetrahedral order,
\begin{equation}
F_\times = -\lambda\Tr[(\bm{\Delta}\hat{\bm{n}})\cdot\bm{O}] = -\lambda\Tr(\bm{\Delta}\cdot\bm{\Omega}),
\label{Ftimes}
\end{equation}
where $\lambda$ is the phenomenological coupling coefficient.  Third, there is the free energy density associated with the presence of tetrahedral order.  In Laudau theory, it can be expressed as an series in powers of the tensor order parameter
\begin{align}
F_\text{tet} &= \frac{1}{2}\mu \Tr(\bm{O}^2) + \frac{1}{4}\nu\Tr(\bm{O}^2)^2
+ \frac{1}{2}\kappa|\bm{\nabla}\bm{O}|^2 \nonumber\\
&= \frac{1}{2}\mu \Tr(\bm{\Omega}^2) + \frac{1}{4}\nu\Tr(\bm{\Omega}^2)^2
+ \frac{1}{2}\kappa|\bm{\nabla}(\bm{\Omega}\hat{\bm{n}})|^2.
\label{Ftet}
\end{align}
The coefficient $\mu$ expresses the quadratic cost or benefit of tetrahedral order; presumably it varies with temperature as $\mu(T)=\mu'(T-T_0)$.  The coefficient $\nu$ keeps the magnitude of tetrahedral order from diverging, and $\kappa$ penalizes gradients of tetrahedral order.

As a \emph{preliminary, over-simplified} calculation, we minimize the total free energy \emph{as if} the splay, twist, bend, and $\bm{\Delta}$ modes were independent of each other.  This minimization gives $S=0$, $T=0$, $\bm{B}=0$, and $\bm{\Delta}=(\lambda/K_{24})\bm{\Omega}$.  By putting those results back into $F$, we obtain the effective free energy density as a function of $\bm{\Omega}$ alone,
\begin{equation}
F_\text{eff} = \frac{1}{2}(\mu-\mu_\text{simple})\Tr(\bm{\Omega}^2) + \frac{1}{4}\nu\Tr(\bm{\Omega}^2)^2
+ \frac{1}{2}\kappa|\bm{\nabla}(\bm{\Omega}\hat{\bm{n}})|^2,
\end{equation}
where the critical value of $\mu$ is
\begin{equation}
\mu_\text{simple}=\frac{\lambda^2}{K_{24}}.
\label{musimple}
\end{equation}
Hence, the minimum depends on whether $\mu$ is above or below $\mu_\text{simple}$.  If $\mu>\mu_\text{simple}$, or equivalently $T>T_\text{simple}=T_0+\lambda^2/(\mu'K_{24})$, the minimum occurs at $\bm{\Omega}=0$ and $\bm{\Delta}=0$.  This state has no tetrahedral order and no director gradients, so it has uniform nematic order, as shown in Fig.~\ref{fig:tetrahedra}(c).  By comparison, if $\mu<\mu_\text{simple}$ or $T<T_\text{simple}$, the minimum occurs at $\bm{\Omega}\not=0$ and $\bm{\Delta}\not=0$, with $\bm{\Delta}=(\lambda/K_{24})\bm{\Omega}$.  That state has nonzero tetrahedral order and nonzero $\bm{\Delta}$ deformation in the director field, and the tetrahedral order and $\bm{\Delta}$ deformation are aligned with each other, as shown in Fig.~\ref{fig:tetrahedra}(d).

We emphasize that the minimization above is \emph{over-simplified}, because the splay, twist, bend, and $\bm{\Delta}$ modes are not independent of each other.  Rather, they must all be derived from the same director field $\hat{\bm{n}}(\bm{r})$.  It is impossible to construct a director field in 3D Euclidean space with a pure, constant, nonzero $\bm{\Delta}$ mode, and no other director deformation modes~\cite{Virga2019}.  Hence, the structure found above can be regarded as an ideal local structure that cannot fill space; i.e.\ it experiences geometric frustration.  To find the optimum global phase, we must do a more complex calculation, in which we minimize the total free energy over director fields that can actually be achieved.  This calculation is done in the following section.

\section{Achievable global phases}
\label{sec:phases}

As discussed in Ref.~\cite{Selinger2022}, a liquid crystal can respond to geometric frustration in two ways.  First, it can fill space with an allowed combination of deformation modes, including the favored mode and some other mode that costs free energy.  Second, it can break space into domains of the favored mode, separated by domain walls or defects.  For a liquid crystal with tetrahedral order, we consider both possibilities.

\subsection{Phases with no defects}
\label{subsec:nodefects}

\begin{figure*}
(a)\includegraphics[height=5.3cm]{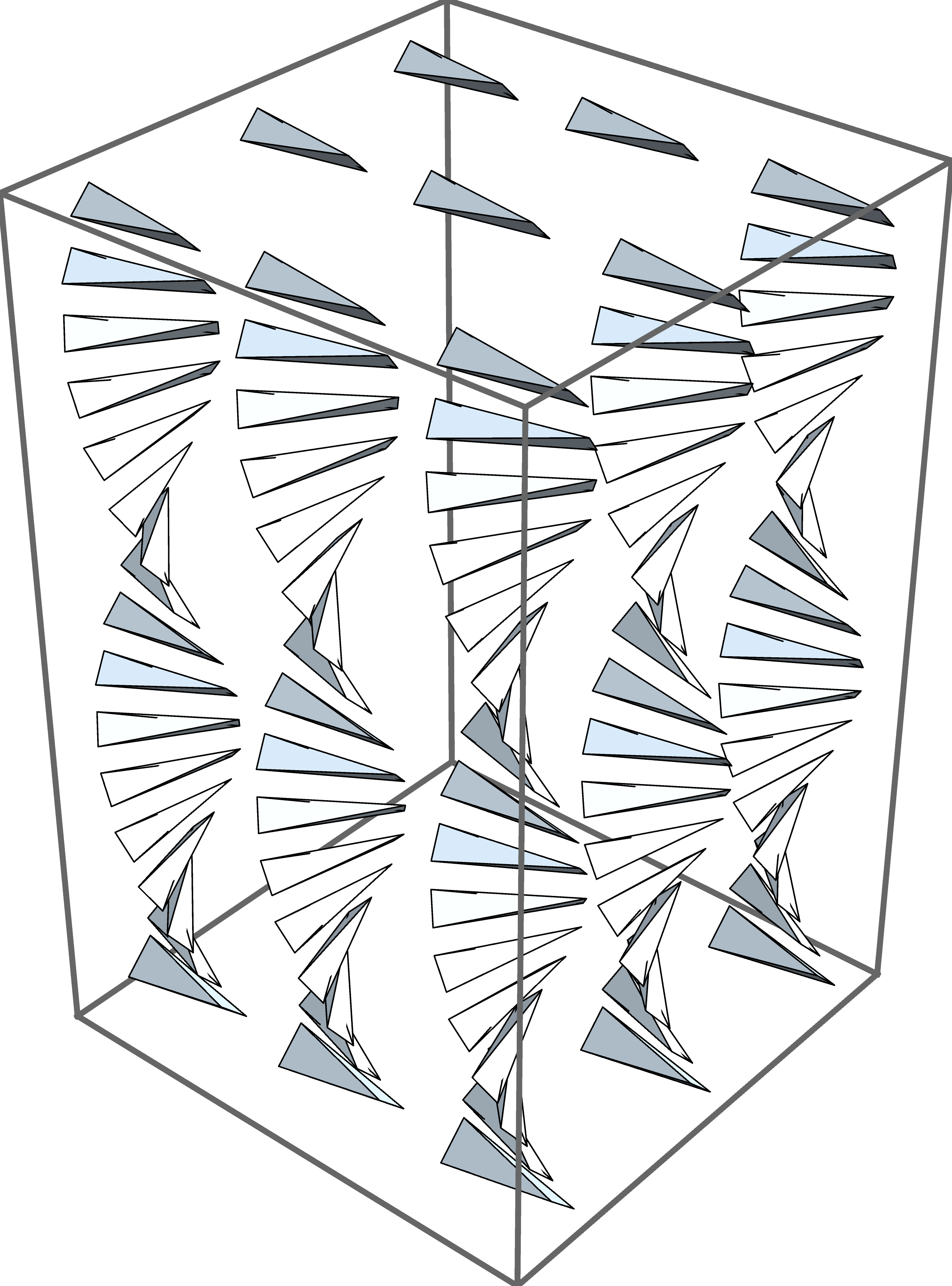}
(b)\includegraphics[height=5.3cm]{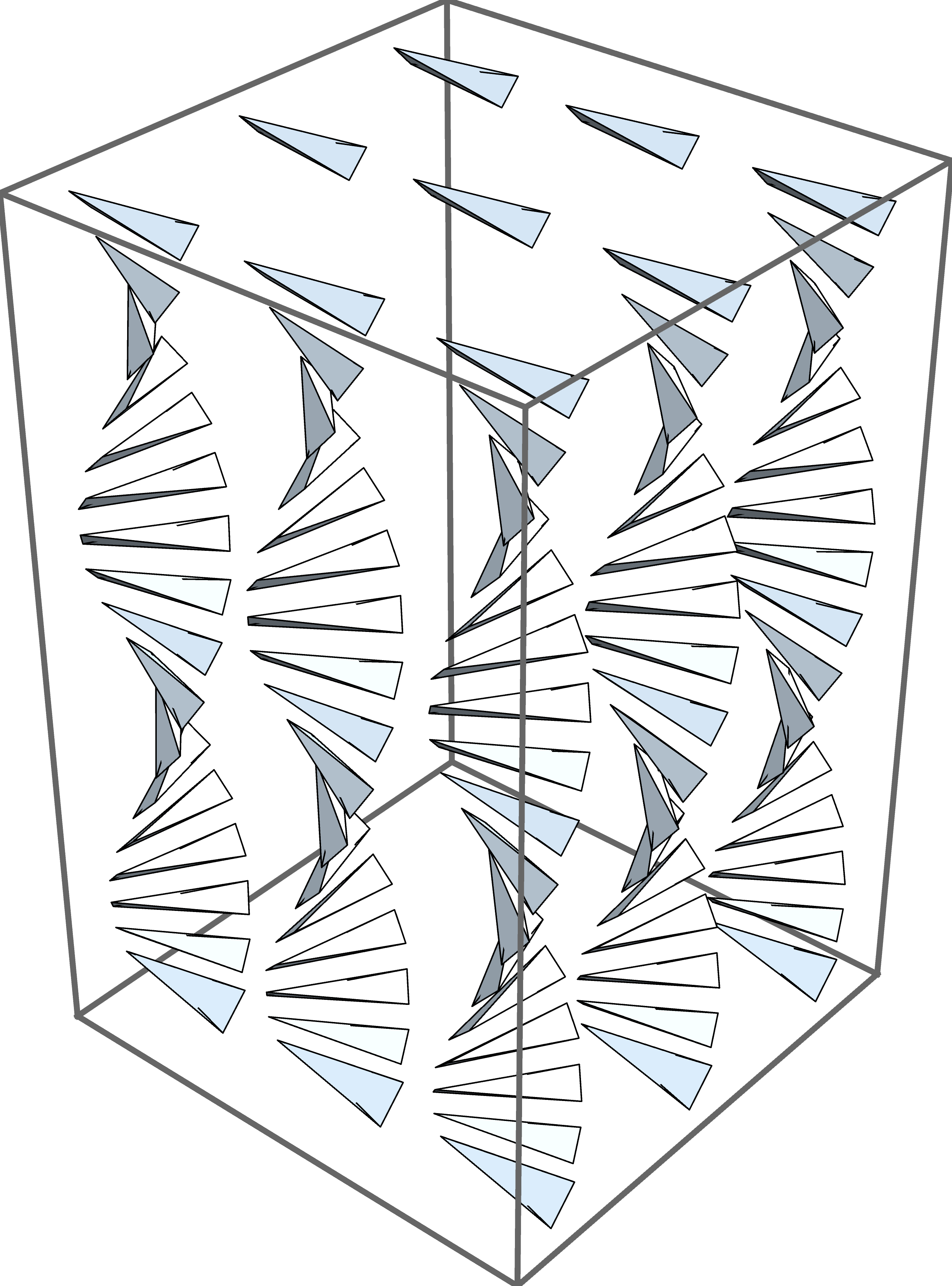}
(c)\includegraphics[height=5.3cm]{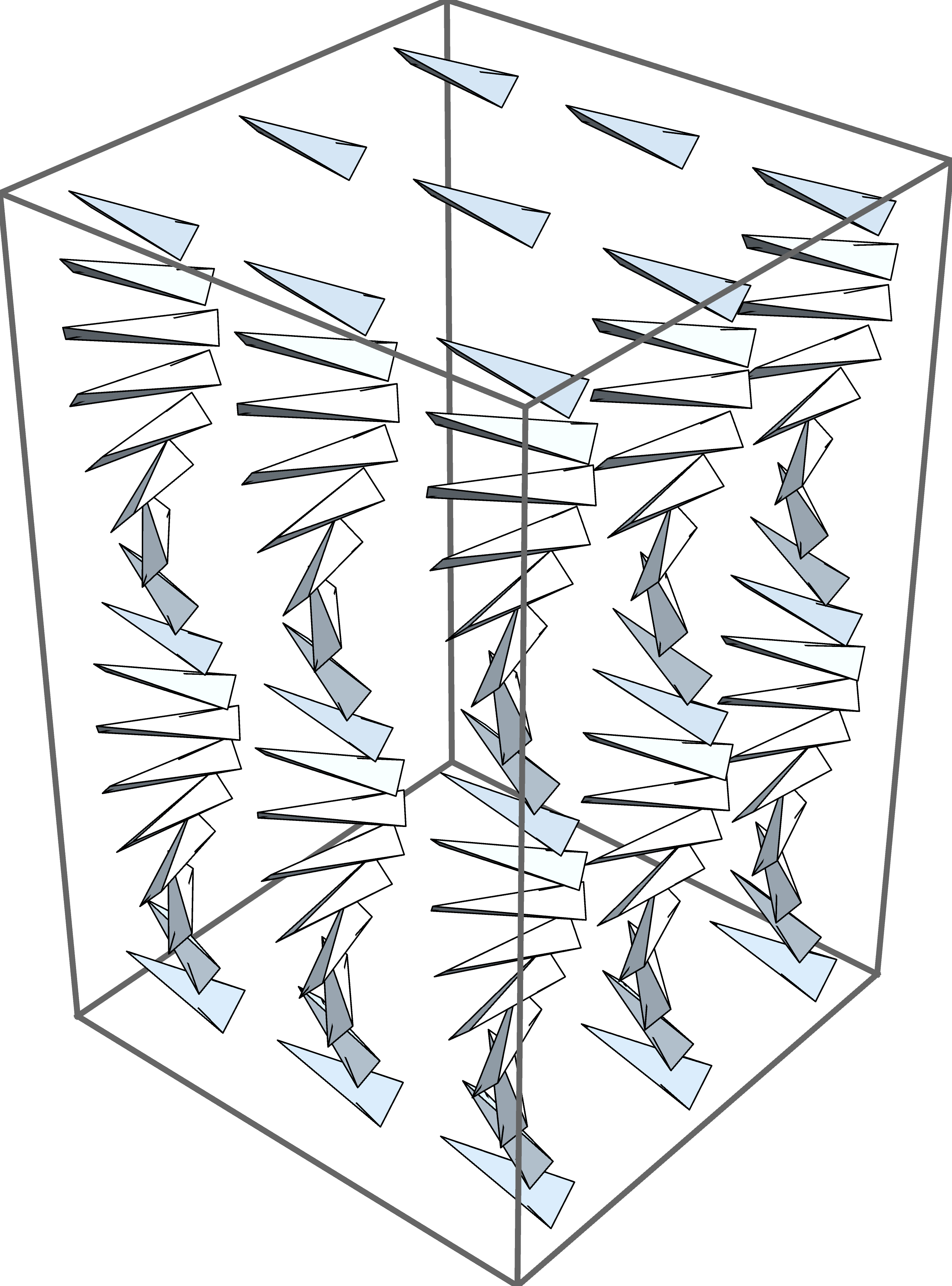}
(d)\includegraphics[height=5.3cm]{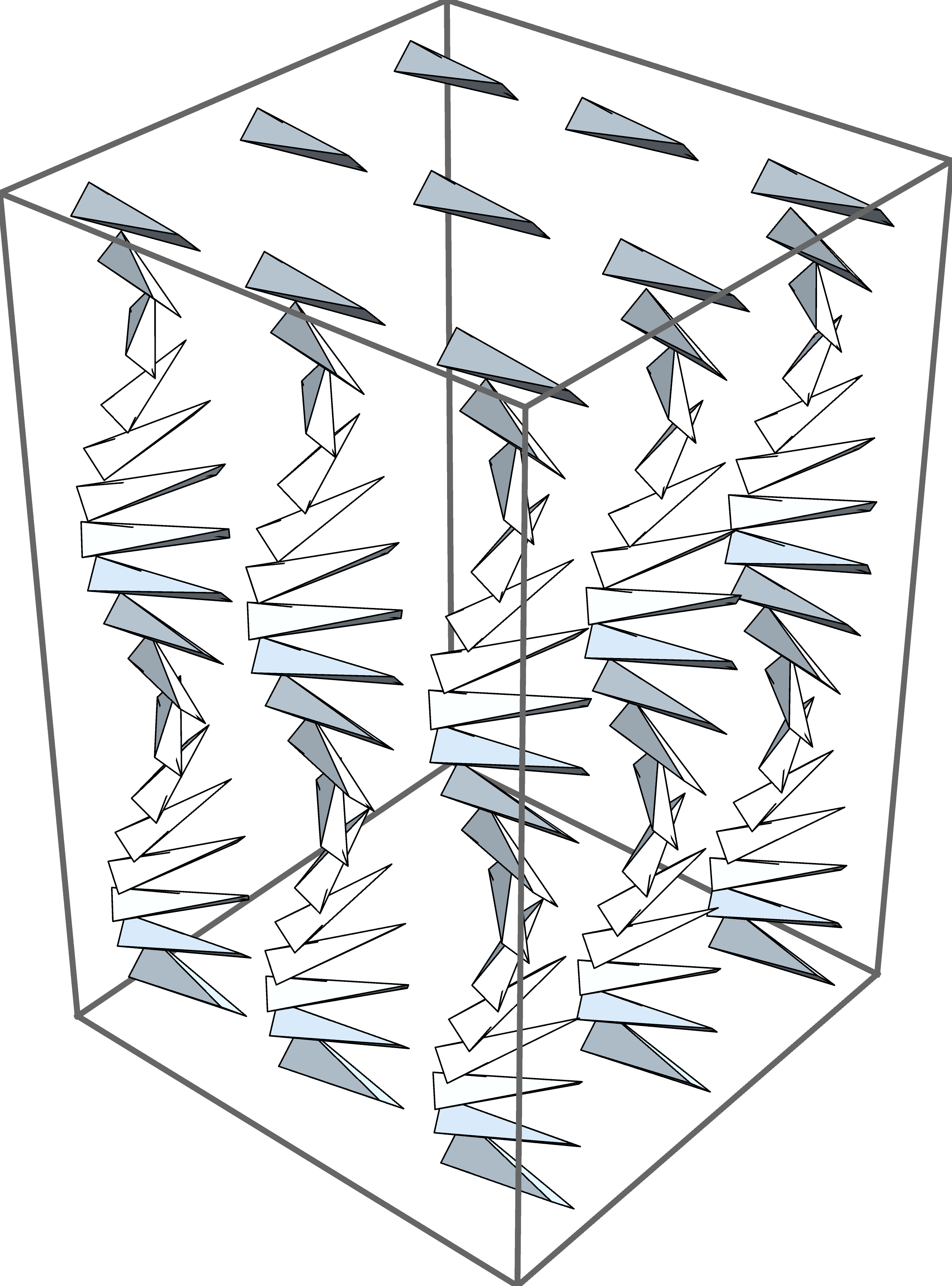}
\caption{Director configuration and tetrahedral order in the spontaneous cholesteric phase.  (a)~$q<0$, $\omega_1<0$.  (b)~$q>0$, $\omega_1>0$.  (c)~$q<0$, $\omega_1>0$.  (d)~$q>0$, $\omega_1<0$.  Structures (a) and (b) are the two degenerate ground states (assuming $\lambda/K_{22}>0$).  Structures (c) and (d) are degenerate with each other, higher in free energy than (a) and (b).}
\label{fig:chol}
\end{figure*}

As proven by Virga~\cite{Virga2019}, there are only two possible combinations of constant deformation modes that can fill 3D Euclidean space.  One is the cholesteric phase, which has a combination of twist and $\bm{\Delta}$ mode.  The other is the twist-bend nematic phase, with a combination of bend, twist, and $\bm{\Delta}$ mode.  We discuss the cholesteric phase in detail, and then briefly show that the twist-bend nematic phase is less favorable than the cholesteric.

\subsubsection{Cholesteric phase}
\label{subsubsec:cholesteric}

A cholesteric phase has a director field with the helical structure illustrated in Fig.~\ref{fig:chol},
\begin{equation}
\hat{\bm{n}} = (\cos qz, \sin qz, 0),
\label{nchol}
\end{equation}
where $q$ is the wavevector.  As shown in Ref.~\cite{Selinger2018}, this cholesteric structure is not pure twist, but rather is a combination of twist and $\bm{\Delta}$ mode.  An explicit calculation gives
\begin{equation}
T=-q,\quad
\bm{\Delta}=\frac{q}{2}
\begin{pmatrix}
0        & 0       & -\sin qz\\
0        & 0       &  \cos qz\\
-\sin qz & \cos qz &        0
\end{pmatrix},
\end{equation}
along with $S=0$ and $\bm{B}=0$.  The eigenvalues of $\bm{\Delta}$ are 0 and $\pm q/2$, and the corresponding eigenvectors are $\hat{\bm{n}}$,
\begin{equation}
\hat{\bm{l}}=\frac{(-\sin qz,\cos qz,1)}{\sqrt{2}}, \quad
\hat{\bm{m}}=\frac{(\sin qz,-\cos qz,1)}{\sqrt{2}}.
\end{equation}

We must now construct an ansatz for the tensor $\bm{\Omega}(z)$ that represents tetrahedral order.  Because $\bm{\Omega}$ is a symmetric, traceless tensor in the plane perpendicular to $\hat{\bm{n}}$, it can be expanded in any basis for that plane.  For example, the eigenvectors $\hat{\bm{l}}$ and $\hat{\bm{m}}$ form a suitable basis.  Hence, we expand $\bm{\Omega}$ as
\begin{align}
\Omega_{ij} &= \omega_1 (l_il_j - m_im_j) + \omega_2 (l_im_j +m_il_j) \nonumber\\
&=\omega_1
\begin{pmatrix}
0        & 0       & -\sin qz\\
0        & 0       &  \cos qz\\
-\sin qz & \cos qz &        0
\end{pmatrix}\nonumber\\
&\quad+\omega_2
\begin{pmatrix}
-\sin^2 qz      & \cos qz \sin qz & 0\\
\cos qz \sin qz & -\cos^2 qz      & 0\\
0               & 0               & 1
\end{pmatrix}.
\label{Omegachol}
\end{align}
Here, the $\omega_1$ term shows the component of tetrahedral order that is aligned with the $\bm{\Delta}$ deformation, because it has the same eigenvectors as $\bm{\Delta}$.  The $\omega_2$ term shows the component of tetrahedral order that has a $45^\circ$ misalignment with the $\bm{\Delta}$ deformation; its eigenvectors are rotated $45^\circ$ from the eigenvectors of $\bm{\Delta}$.

We insert Eqs.~(\ref{nchol}) and~(\ref{Omegachol}) for the cholesteric director and tetrahedral order into Eqs.~(\ref{Fnem}--\ref{Ftet}) for the free energy density, and obtain
\begin{align}
F_\text{chol} &= \frac{1}{2} K_{22} q^2 - \lambda q \omega_1 + \mu (\omega_1^2 + \omega_2^2)
+ \nu (\omega_1^2 + \omega_2^2)^2 \nonumber \\
&\quad + 2 \kappa q^2 (\omega_1^2 + \omega_2^2).
\end{align}
Minimizing over the variational parameters $q$ and $\omega_2$ gives $q=\lambda\omega_1/(K_{22}+4\kappa\omega_1^2)$ and $\omega_2=0$, so that there is no component of tetrahedral order that is misaligned by $45^\circ$ from $\bm{\Delta}$.  The effective free energy density can then be expanded as a power series in $\omega_1$,
\begin{equation}
F_\text{chol} = \left(\mu - \frac{\lambda^2}{2 K_{22}} \right)\omega_1^2
+ \left(\nu + \frac{2\kappa\lambda^2}{K_{22}^2}\right)\omega_1^4.
\end{equation}
From the coefficient of the quadratic term, we see that the critical point occurs at
\begin{equation}
\mu_\text{chol}=\frac{\lambda^2}{2 K_{22}},
\label{muchol}
\end{equation}
or equivalently at temperature $T_c=T_0+\lambda^2/(\mu'K_{22})$.  Above the critical point, the minimum has $\omega_1=0$ and $q=0$, and hence the liquid crystal is in a uniform nematic phase.  Below the critical point, the minimum can be expanded as
\begin{align}
\omega_1 &=\pm\left[\frac{\mu_\text{chol}-\mu}{2(\nu+2\kappa\lambda^2/K_{22}^2)}\right]^{1/2},\nonumber\\
q &=\frac{\lambda\omega_1}{K_{22}}
=\pm\frac{\lambda}{K_{22}}\left[\frac{\mu_\text{chol}-\mu}{2(\nu+2\kappa\lambda^2/K_{22}^2)}\right]^{1/2},
\end{align}
and hence the liquid crystal has tetrahedral order and a cholesteric helix.  The free energy density of this ordered phase becomes
\begin{equation}
F_\text{chol}=-\frac{(\mu_\text{chol}-\mu)^2}{4(\nu+2\kappa\lambda^2/K_{22}^2)}.
\label{Fchol}
\end{equation}

Note that the ordered phase has two solutions with positive and negative $q$, corresponding to right- and left-handed helices, and these two solutions have equal free energy.  Hence, the liquid crystal has a spontaneous symmetry breaking from a high-temperature achiral state to a low-temperature chiral state, which is equally likely to be right- or left-handed.  We might describe the ordered state as a spontaneous cholesteric phase, in contrast to a conventional cholesteric phase of chiral molecules, which has a single preferred handedness.  Note also that the parameter $\omega_1$ has a sign consistent with $q$, so that the tetrahedral order is compatible with the handedness of the cholesteric helix.

To illustrate this point, Figs.~\ref{fig:chol}(a) and~(b) show mirror-image structures, (a)~with $q$ and $\omega_1$ both negative, (b)~with $q$ and $\omega_1$ both positive.  These structures are the two ground states (assuming $\lambda/K_{22}>0$), with equal free energy.  By contrast, Figs.~\ref{fig:chol}(c) and~(d) show structures with $q$ negative and $\omega_1$ positive, or vice versa.  Those structures are equal in free energy to each other, but higher in free energy than (a) and (b).  We can see that (a) and (b) have more efficient packings of the tetrahedral particles than (c) and (d).

One might ask why the results of this calculation depend on the elastic constant $K_{22}$ rather than on $K_{24}$, as in the over-simplified theory of Sec.~\ref{sec:tetrahedra}.  We have both a mathematical explanation and a physical explanation.  The mathematical explanation is that the coefficient $K_{24}$ multiplies the combination of terms $[\Tr(\bm{\Delta}^2)-\frac{1}{2}S^2-\frac{1}{2}T^2]$ in the Oseen-Frank free energy density~(\ref{Fnem}).  As discussed in Ref.~\cite{Selinger2018}, this combination is a total divergence, and hence its volume integral can be transformed into a surface integral.  For that reason, this combination does not contribute to the free energy of any periodic structure.  Thus, its coefficient $K_{24}$ cannot enter any results.

The physical explanation is that, in our formalism~\cite{Selinger2018}, $K_{22}$ is actually the sum of two more fundamental elastic constants:  $K_{24}$ for the $\bm{\Delta}$ mode and $(K_{22}-K_{24})$ for pure twist (which is double twist).  The spontaneous cholesteric phase includes both the favorable $\bm{\Delta}$ deformation and the unfavorable twist deformation, and hence its energy cost involves the sum of those two elastic constants.  For a typical liquid crystal that satisfies the Ericksen inequality $K_{22}>K_{24}$~\cite{Ericksen1966,Selinger2018}, the critical point $\mu_\text{chol}$ of Eq.~(\ref{muchol}) is lower than the over-simplified $\mu_\text{simple}$ of Eq.~(\ref{musimple}).  Hence, geometric frustration makes it more difficult for the liquid crystal to go into a nonuniform tetrahedral phase, compared to the over-simplified theory with no geometric frustration.  (Certain lyotropic chromonic liquid crystals have $K_{22}<K_{24}$~\cite{Davidson2015}, violating the Ericksen inequality.  In those materials, there is a spontaneous tendency to twist, which would make it easier for the liquid crystal to go into a nonuniform tetrahedral phase.)

\subsubsection{Twist-bend nematic phase}
\label{subsubsec:ntb}

Apart from the cholesteric phase, the other possible combination of constant deformation modes is the twist-bend nematic ($N_{TB}$) phase.  In the $N_{TB}$ phase, the director varies in a helix, but it is not perpendicular to the helical axis.  Rather, the director maintains a constant cone angle $\beta$ with respect to the helical axis.  For that reason, the $N_{TB}$ phase is sometimes described as a ``heliconical'' structure.

Many theoretical and experimental studies~\cite{Meyer1976,Dozov2001,Chen2013,Borshch2013,Shamid2013,Barbero2015} have shown that the $N_{TB}$ phase may occur in systems of bent-core molecules, which tend to form polar order perpendicular to the director and hence have a favored bend.  Here, we consider whether the $N_{TB}$ phase can occur in systems of distorted tetrahedra as in Fig.~\ref{fig:tetrahedra}, which tend to form tetrahedral order and hence have a favored $\bm{\Delta}$ mode.

For this calculation, we begin with the director field of an $N_{TB}$ phase,
\begin{equation}
\hat{\bm{n}} = (\sin\beta\cos qz, \sin\beta\sin qz, \cos\beta).
\end{equation}
An explicit calculation shows that this director field has bend, twist, and $\bm{\Delta}$ deformations, but no splay.  We proceed just as in the cholesteric case above:  derive the $\hat{\bm{l}}$ and $\hat{\bm{m}}$ vectors, construct the tetrahedral order tensor $\bm{\Omega}$, put the director field and order tensor into the free energy density, and minimize over variational parameters.  The calculation shows that the angle $\beta$ is always driven to $\pi/2$, so that the $N_{TB}$ phase becomes the spontaneous cholesteric phase.  For this model, the spontaneous cholesteric phase always has a lower free energy than the $N_{TB}$ phase with $\beta\not=\pi/2$.  Hence, we will not consider the $N_{TB}$ phase further in this study.

\subsection{Phases with defects}
\label{subsec:defects}

\begin{figure*}
(a)\includegraphics[height=10cm]{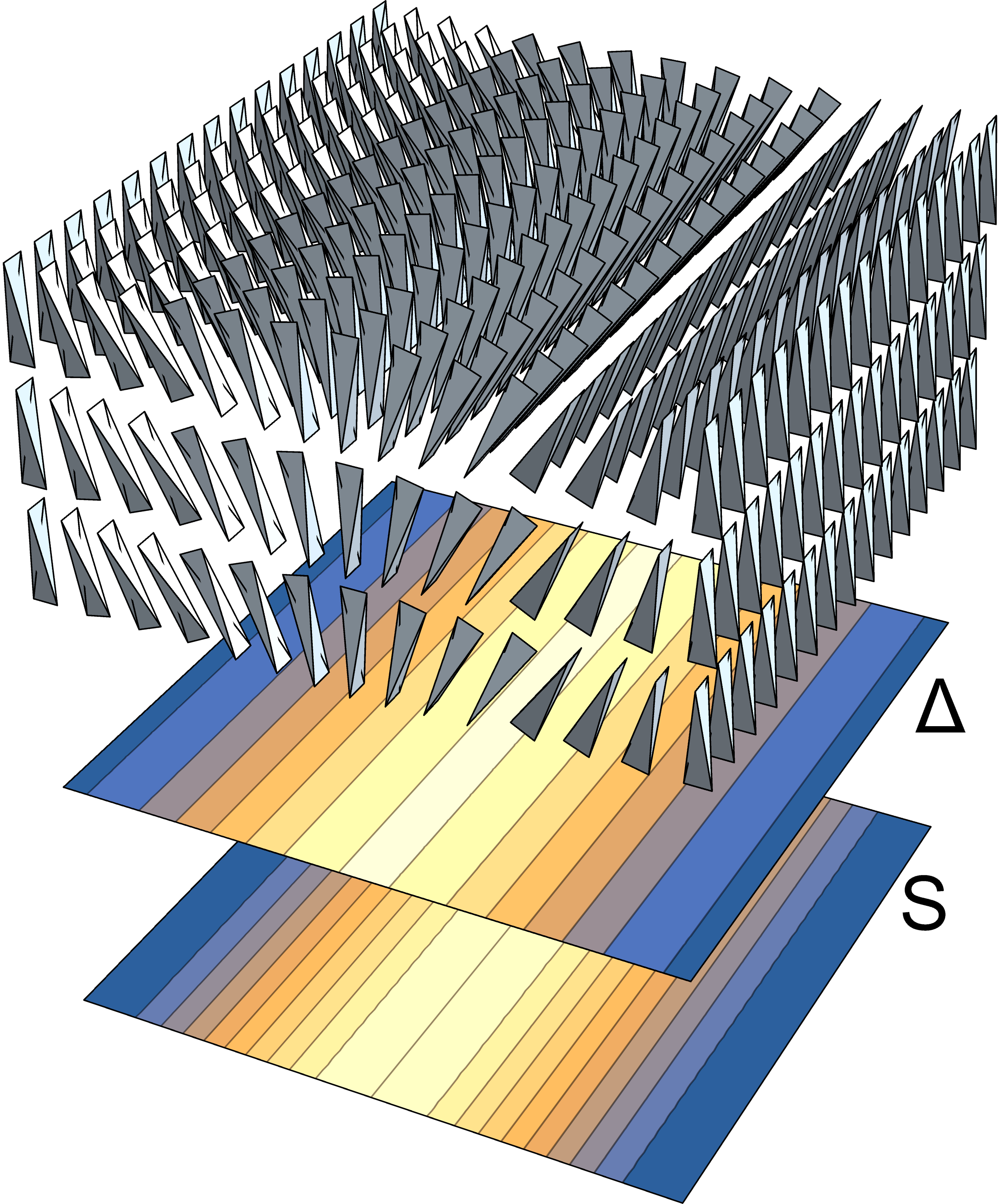}
(b)\includegraphics[height=10cm]{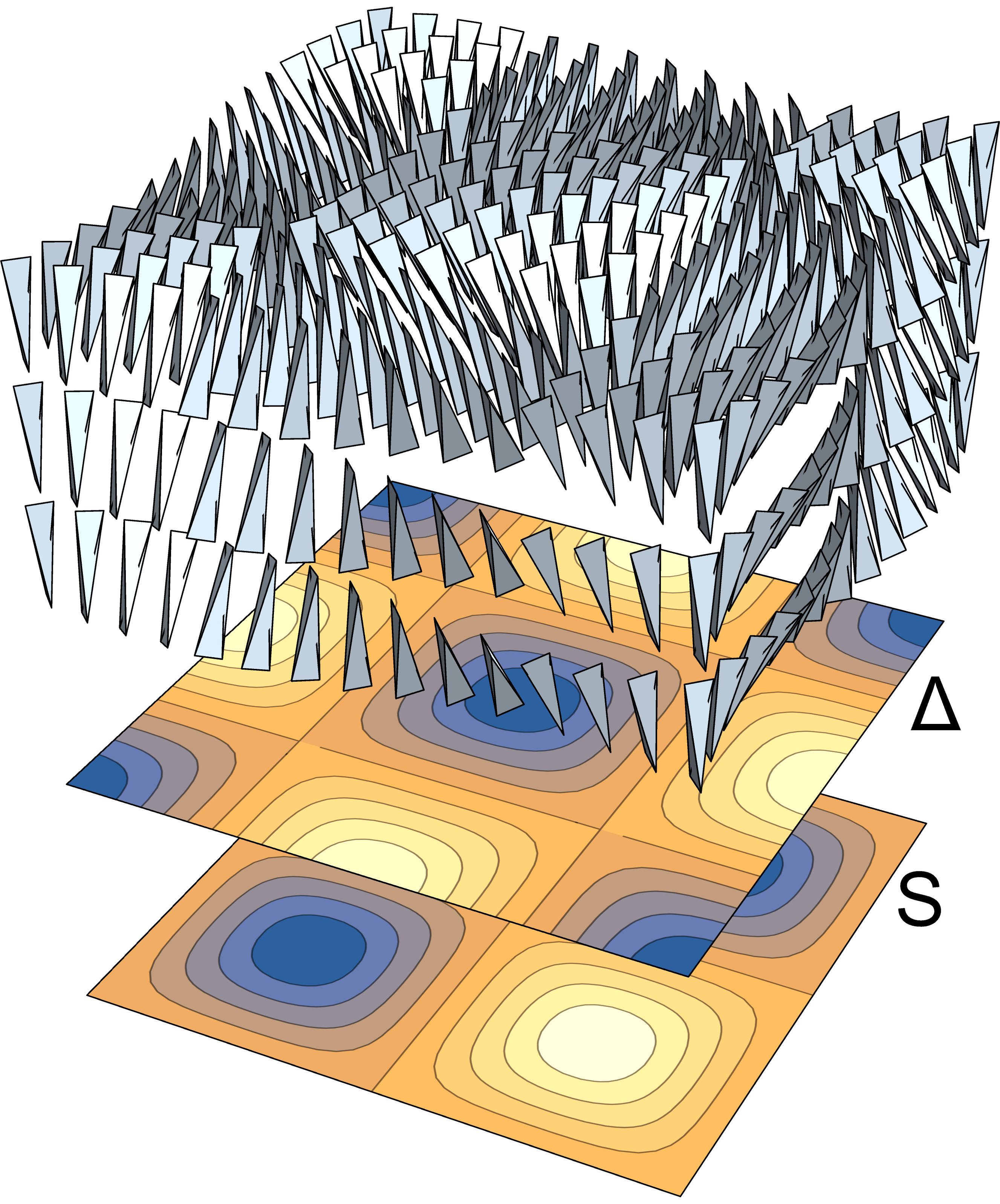}
\caption{Director configuration and tetrahedral order in the splay nematic phases.  (a)~Single splay.  (b)~Double splay.  Below the structures, the upper contour plot shows one of the nonzero eigenvalues of the $\bm{\Delta}$ deformation tensor, and the lower contour plot shows the splay.  In both contour plots, bright (yellow) represents positive, and dark (blue) represents negative.}
\label{fig:splaynematic}
\end{figure*}

The second way that a liquid crystal can respond to geometric frustration is to break space into domains of approximately the ideal local structure, separated by domain walls, which can be regarded as defects in the ideal local structure.  We would like to determine whether this type of domain structure can occur in systems that form tetrahedral order and a favored $\bm{\Delta}$ deformation.

Of course, we do not have a complete list of possible domain structures.  However, we already know two domain structures that include a large component of $\bm{\Delta}$ deformation, which have previously been proposed as models for the splay nematic phase, induced by polar order parallel to the director.  One of these structures, which we call single splay, was proposed by the experimental group that reported the splay nematic phase~\cite{Mertelj2018}.  The other structure, which we call double splay, is a more complex alternative that we investigated in a previous article~\cite{Rosseto2020}.  Here, we consider whether these structures can be induced by tetrahedral order and the favored $\bm{\Delta}$ mode.

\subsubsection{Single splay}
\label{subsubsec:single}

The single-splay phase has the structure proposed in Ref.~\cite{Mertelj2018}, which is illustrated in Fig.~\ref{fig:splaynematic}(a).  The director field varies in the $(x,z)$ plane as a function of $x$.  It has alternating domains of upward and downward splay, together with $\bm{\Delta}$ mode, separated by walls that are mainly bend.  The director field can be written as
\begin{equation}
\hat{\bm{n}}=(\sin\theta(x), 0, \cos\theta(x)),\quad\theta(x)=\theta_0\sin q x,
\label{nss}
\end{equation}
with $\theta_0$ small.  The four director deformation modes can be calculated explicitly as
\begin{align}
&S=q\theta_0 \cos q x \cos\theta(x), \quad T=0,\nonumber\\
&\bm{B}=\frac{1}{2}q\theta_0 \cos q x (-\sin2\theta(x),0,1-\cos2\theta(x)),\nonumber\\
&\bm{\Delta}=\frac{1}{4}S
\begin{pmatrix}
1+\cos2\theta(x)              & 0  & -\sin2\theta(x)\\
0                               & -2 & 0\\
-\sin2\theta(x) & 0  & 1-\cos2\theta(x)
\end{pmatrix}.
\end{align}
The eigenvalues of $\bm{\Delta}$ are 0 and $\pm S/2$, and the corresponding eigenvectors are $\hat{\bm{n}}$,
\begin{equation}
\hat{\bm{l}}=(-\cos\theta(x), 0, \sin\theta(x)), \quad\hat{\bm{m}}=(0,1,0).
\end{equation}
To construct an ansatz for the tetrahedral order, we use
\begin{equation}
\Omega_{ij}(x)=(\omega \cos q x)(l_i l_j - m_i m_j),
\label{Omegass}
\end{equation}
so that $\bm{\Omega}$ has the same eigenvectors as $\bm{\Delta}$, and the eigenvalues depend on position in approximately the same way.  This ansatz is illustrated by the distorted tetrahedra in Fig.~\ref{fig:splaynematic}(a).  It is quite compatible with the splay deformation, and gives an efficient packing of the tetrahedral particles.

We insert Eqs.~(\ref{nss}) and~(\ref{Omegass}) for the director and tetrahedral order into Eqs.~(\ref{Fnem}--\ref{Ftet}) for the free energy density, expand in a power series up to fourth order in $\theta_0$, and then average over position.  The average free energy density becomes
\begin{align}
F_\text{ss}&=
\frac{K_{11}q^2\theta_0^2}{4}\left[1-\frac{\theta_0^2}{4}\right]
+\frac{K_{33}q^2\theta_0^4}{16}
-\frac{\lambda q\theta_0\omega}{2}\left[1-\frac{\theta_0^2}{8}\right]\nonumber\\
&\quad+\frac{\mu\omega^2}{2}
+\frac{3\nu\omega^4}{8}
+ \frac{\kappa q^2\omega^2}{2}\left[1+\frac{3\theta_0^2}{2}\right],
\end{align}
with the subscript ss for single splay.  By minimizing over the variational parameters $q$, $\theta_0$, and $\omega$, we find that the uniform nematic state with $\theta_0=0$ and $\omega=0$ becomes unstable at the critical point
\begin{equation}
\mu_\text{ss}=\frac{\lambda^2}{2K_{11}}.
\end{equation}
Below that critical point, the variational parameters scale as
\begin{align}
&q=\left[\frac{\mu_\text{ss}-\mu}{3\kappa}\right]^{1/2},\quad
\theta_0=\frac{2K_{11}}{\lambda}\left[\frac{2(\mu_\text{ss}-\mu)}{3K_{33}}\right]^{1/2},\nonumber\\
&\omega = \frac{2K_{11}^2(\mu_\text{ss}-\mu)}{3\lambda^2}\left[\frac{2}{K_{33}\kappa}\right]^{1/2},
\end{align}
and the average free energy density (relative to the uniform nematic phase) scales as
\begin{equation}
F_\text{ss}=-\frac{4K_{11}^4(\mu_\text{ss}-\mu)^3}{27K_{33}\kappa\lambda^4}.
\end{equation}

We note that these results depend on the elastic constant $K_{11}$ rather than $K_{24}$, for the same mathematical and physical reasons that the results of Sec.~\ref{subsubsec:cholesteric} depend on $K_{22}$ rather than $K_{24}$.

Also, we note that these results are very similar to previous results for the same structure driven by a different mechanism:  not tetrahedral order, but polar order parallel to the director~\cite{Mertelj2018,Rosseto2020}.

\subsubsection{Double splay}
\label{subsubsec:double}

The double-splay phase is an alternative structure proposed in Ref.~\cite{Rosseto2020}, which has the director configuration shown in Fig.~\ref{fig:splaynematic}(b).  It has symmetry of a checkerboard, with alternating positive and negative splay in each square.  Unlike the single-splay phase, the $\bm{\Delta}$ mode is not concentrated in the same location as the splay; rather, $\bm{\Delta}$ mode is concentrated near the corners where four splay squares come together.  The director field can be described mathematically as
\begin{equation}
\hat{\bm{n}}=
\frac{(\theta_0\sin qx\cos qy,\theta_0\sin qy\cos qx,1)}{\left[1+\theta_0^2 (\sin^2 qx\cos^2 qy+\sin^2 qy\cos^2 qx)
\right]^{1/2}},
\end{equation}
again with $\theta_0$ small.  Following the same procedure as the previous sections, we calculate the four deformation modes, and we diagonalize $\bm{\Delta}$ to find the eigenvalues and eigenvectors $\hat{\bm{n}}$, $\hat{\bm{l}}$, and $\hat{\bm{m}}$.  The calculations are much longer than in the previous sections, and can only be done as power series expansions in $\theta_0$; we omit the results to save space.  As an ansatz for the tetrahedral order, we use
\begin{equation}
\Omega_{ij}(x,y)=(\omega \sin qx\sin qy)(l_i l_j - m_i m_j),
\end{equation}
again so that $\bm{\Omega}$ has the same eigenvectors as $\bm{\Delta}$, and the eigenvalues depend on position in approximately the same way.  This ansatz is shown by the distorted tetrahedra in Fig.~\ref{fig:splaynematic}(b), and it gives an efficient packing of these particles in the director field with double splay.

By inserting these assumptions into the free energy density and averaging over the $x$ and $y$ periodicity, we obtain
\begin{align}
F_\text{ds}&=
\frac{K_{11}q^2\theta_0^2}{2}\left[1-\frac{5\theta_0^2}{8}\right] 
+\frac{K_{33}q^2\theta_0^4}{16}
-\frac{\lambda q\theta_0\omega}{2}\left[1-\frac{5\theta_0^2}{16}\right]\nonumber\\
&+\frac{\mu\omega^2}{4}
+\frac{9\nu\omega^4}{64}
+\frac{\kappa q^2\omega^2}{2}\left[1+\frac{5\theta_0^2}{4}-\frac{45\theta_0^4}{64}\right],
\end{align}
with the subscript ds for double splay.  We minimize that free energy density over the variational parameters $q$, $\theta_0$, and $\omega$, and find that the uniform nematic state becomes unstable at the critical point
\begin{equation}
\mu_\text{ds}=\frac{\lambda^2}{2K_{11}}.
\end{equation}
which is exactly the same as $\mu_\text{ss}$ for the single-splay phase transition.  In the double-splay phase below $\mu_\text{ds}$, the variational parameters scale as
\begin{align}
&q=\left[\frac{\mu_\text{ds}-\mu}{6\kappa}\right]^{1/2},\quad
\theta_0=\frac{4K_{11}}{\lambda}\left[\frac{\mu_\text{ds}-\mu}{3 \,K_{33}}\right]^{1/2},\nonumber\\
&\omega = \frac{4K_{11}^2(\mu_\text{ds}-\mu)}{3\lambda^2}\left[\frac{2}{K_{33}\kappa}\right]^{1/2},
\end{align}
and the average free energy density as
\begin{equation}
F_\text{ds}=-\frac{8K_{11}^4(\mu_\text{ds}-\mu)^3}{27K_{33}\kappa\lambda^4}.
\label{Fds}
\end{equation}
Comparing $F_\text{ds}$ and $F_\text{ss}$, we see that the free energy of the double-splay phase is twice as negative as the free-energy of the single-splay phase, with both measured relative to the uniform nematic phase.  Hence, the double-state phase is always energetically preferred over the single-splay phase, at least in the regime close to the critical point where this calculation is valid.

\subsection{Phase diagram}
\label{subsec:phasediagram}

\begin{figure}
\includegraphics[width=\columnwidth]{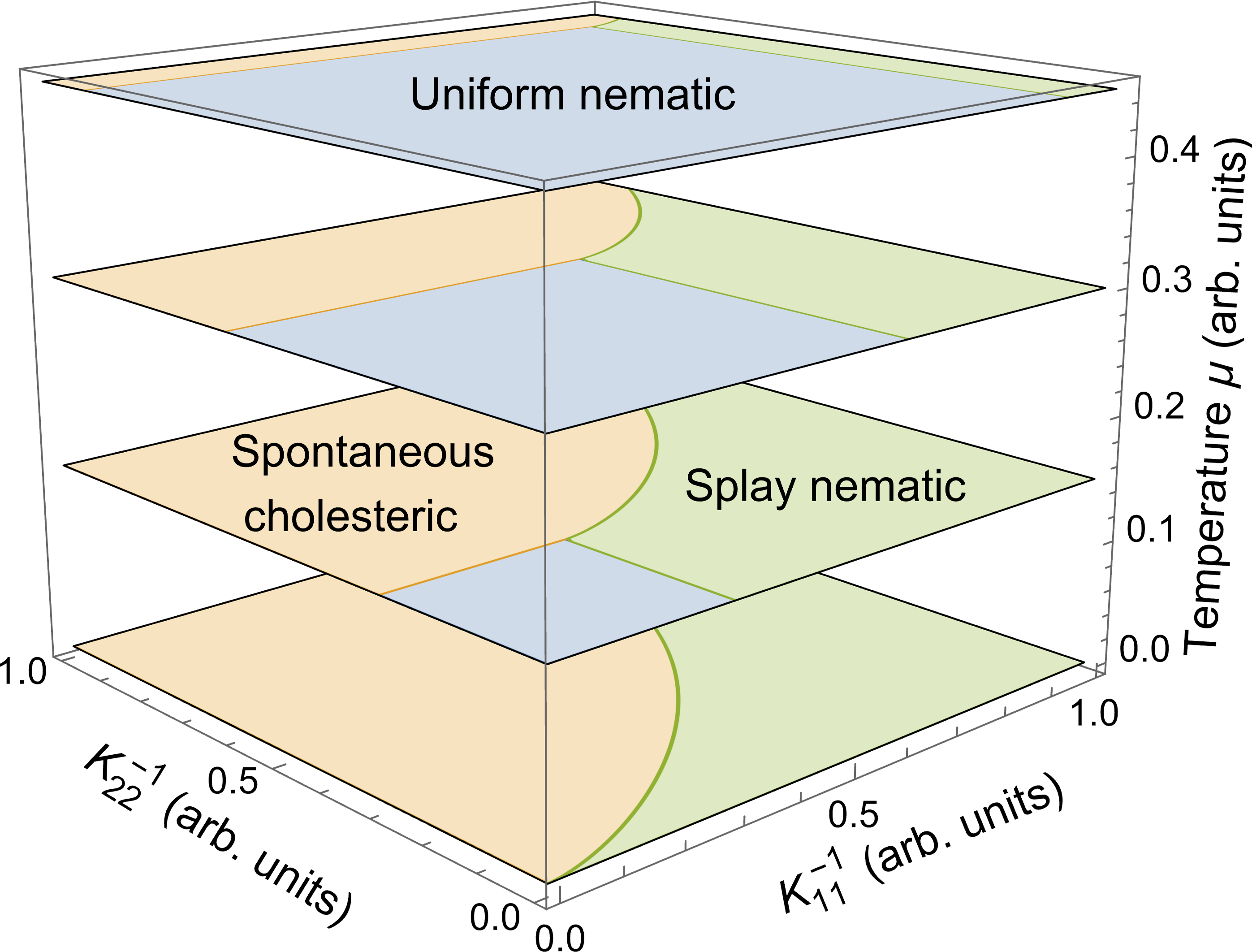}
\caption{Schematic phase diagram in terms of elastic constants $K_{11}^{-1}$, $K_{22}^{-1}$, and the temperature-like variable $\mu=\mu'(T-T_0)$, all in arbitrary units.}
\label{fig:phasediagram}
\end{figure}

The results of this section are summarized in the phase diagram of Fig.~\ref{fig:phasediagram}.  Here, the axes are the inverse elastic constants $1/K_{11}$ and $1/K_{22}$ and the temperature-like variable $\mu=\mu'(T-T_0)$.  All three variables are expressed in arbitrary units such that $\lambda=\nu=\kappa=1$.  At high temperature, when $\mu$ is large and positive, there is a high free energy penalty for the system to develop tetrahedral order.  In that regime, the system is in a uniform nematic phase.  As $\mu$ decreases toward zero, it becomes easier for the system to develop tetrahedral order, accompanied by the $\bm{\Delta}$ director deformation.  When $\mu$ reaches zero, there is no longer any penalty for tetrahedral order, and hence the uniform nematic phase disappears.

The configuration of the tetrahedral order and the director field depends on $K_{11}$ and $K_{22}$.  If $K_{22}$ is small and $K_{11}$ is large, the system develops tetrahedral order and the $\bm{\Delta}$ mode together with $\emph{twist}$, and hence it forms the spontaneous cholesteric phase.  By contrast, if $K_{11}$ is small and $K_{22}$ is large, the system develops tetrahedral order and the $\bm{\Delta}$ mode together with $\emph{splay}$, and hence forms the splay nematic phase, preferably with the double splay structure.

If $\mu$ is below both $\mu_\text{chol}=\lambda^2/(2K_{22})$ and $\mu_\text{ds}=\lambda^2/(2K_{11})$, the system might be either in the spontaneous cholesteric or the splay nematic phase.  To determine which phase is favored, we compare the free energies of Eqs.~(\ref{Fchol}) and (\ref{Fds}).  This construction gives the curved phase boundary shown in Fig.~\ref{fig:phasediagram}.  We recognize that these equations for the free energies are only valid in the regimes close to the critical points, and hence the phase boundary is only a rough approximation.  A better phase boundary would require a more detailed model for the spontaneous cholesteric and splay nematic phases, which is beyond the scope of this article.

\section{Discussion}
\label{sec:discussion}

In this article, we have developed a Landau theory for tetrahedral order in nematic liquid crystals.  In the ideal local structure, tetrahedral order is coupled with the $\bm{\Delta}$ mode of director deformation.  Because of geometric frustration, the $\bm{\Delta}$ mode cannot fill space by itself.  Rather, it must be accompanied by a certain amount of twist or splay.  If $K_{22}<K_{11}$, then twist has a lower free energy penalty than splay, and hence tetrahedral order leads to the formation of a spontaneous cholesteric phase.  By contrast, if $K_{11}>K_{22}$, then splay has a lower free energy penalty than twist, so that tetrahedral order induces the formation of a splay nematic phase.  Thus, the theory gives the phase diagram of Fig.~\ref{fig:phasediagram}.

The predictions of this theory might be tested in simulations of particles with the distorted tetrahedral shape shown in Fig.~\ref{fig:tetrahedra}(b).  Based on packing considerations, these particles would have a natural tendency toward tetrahedral order, together with the $\bm{\Delta}$ deformation of the long axes.  If this tendency is strong enough, we would expect simulations to show the spontaneous cholesteric or splay nematic phases predicted here.  Ideally, the predictions might also be realized in molecular systems with the appropriate chemical structure.  We do not see any immediate prospect for designing such molecules, but it is a theoretical possibility.

Apart from the specific phase diagram predicted here, the theory has two general implications for studies of director deformations and modulated phases in liquid crystals.

First, the theory demonstrates the importance of the $\bm{\Delta}$ mode of director deformation.  Some researchers believe that the $\bm{\Delta}$ mode can be ignored, just because the elastic constant $K_{24}$ multiplies a total divergence in the Oseen-Frank free energy density.  That is a misconception.  Our review article~\cite{Selinger2018} already argued that many classic phenomena in liquid crystals can be understood most easily in terms of $\bm{\Delta}$ as a bulk elastic mode, on a par with splay, twist, and bend.  The current article shows theoretically that a liquid crystal can have a favored $\bm{\Delta}$ mode, just as chiral liquid crystals have a favored twist and polar liquid crystals have a favored bend or splay.  The favored $\bm{\Delta}$ mode controls the bulk phase diagram, in spite of the fact that $K_{24}$ does not appear in the predictions.  This result is possible because the predictions involve the sum of $K_{24}$ with the elastic constant $(K_{22}-K_{24})$ for pure double twist, or the elastic constant $(K_{11}-K_{24})$ for pure double splay.

Second, the theory shows that it can be difficult to relate a liquid crystal's director configuration to the underlying physical mechanism that caused that configuration.  A cholesteric phase is normally caused by favored twist in a chiral liquid crystal, but the theory predicts that it can also be caused by a favored $\bm{\Delta}$ mode.  Likewise, a splay nematic phase may be caused by favored splay in a polar liquid crystal, but the theory shows that it can also be caused by a favored $\bm{\Delta}$ mode.  Our review article~\cite{Selinger2022} argued that a favored $\bm{\Delta}$ mode can masquerade as chirality in smectic liquid crystals; here we see that the masquerade can also occur in nematic liquid crystals and with splay.  Hence, one cannot just observe the director configuration and infer that a liquid crystal has a favored twist or splay.  Rather, one must also consider the shape of the constituent molecules, and develop a more detailed model of how that shape is related to the director configuration.  This result provides a certain note of caution for future research.

\acknowledgments

This work was supported in part by National Science Foundation Grant DMR-1409658 and by Coordena\c{c}\~ao de Aperfei\c{c}oamento de Pessoal de N\'ivel Superior - Brazil (CAPES).

\bibliography{ref}

\end{document}